\documentclass{natureASTRO}

\pdfoutput=1

\usepackage{graphicx}
\usepackage{color}
\usepackage{hyperref}
\usepackage{longtable}
\usepackage{lineno}

\usepackage{pdflscape}
\usepackage{amsmath, amssymb}
\usepackage{lscape}

\title{A supra-massive population of stellar-mass black holes in the globular cluster Palomar 5}

\author{M. Gieles$^{1, 2}$,  D. Erkal$^{3}$, F. Antonini$^{4}$, Eduardo Balbinot$^5$ and Jorge Pe\~{n}arrubia$^{6,7}$}

\begin{document}

\maketitle

\begin{affiliations}
\item{ICREA, Pg. Llu\'{i}s Companys 23, E08010 Barcelona, Spain}
\item{Institut de Ci\`{e}ncies del Cosmos (ICCUB), Universitat de Barcelona (IEEC-UB), Mart\'{i} i Franqu\`{e}s 1, E08028 Barcelona, Spain}
\item{Department of Physics, University of Surrey, Guildford, GU2 7XH, Surrey, UK}
\item{Gravity Exploration Institute, School of Physics and Astronomy, Cardiff University, Cardiff, CF24 3AA, UK}
\item{Kapteyn Astronomical Institute, University of Groningen, Postbus 800, NL-9700AV Groningen, The Netherlands}
\item{Institute for Astronomy, University of Edinburgh, Royal Observatory, Blackford Hill, Edinburgh EH9 3HJ, UK}
\item{Centre for Statistics, University of Edinburgh, School of Mathematics, Edinburgh EH9 3FD, UK}\\
\end{affiliations}

\begin{abstract}
Palomar 5 is one of the sparsest star clusters in the Galactic halo and is best-known for its spectacular tidal tails, spanning over 20 degrees across the sky. With $N$-body simulations we show that both distinguishing features can result from a stellar-mass black hole population, comprising $\sim20\%$ of the present-day cluster mass. In this scenario, Palomar 5 formed with a `normal' black hole mass fraction of a few per cent, but stars were lost at a higher rate than black holes, such that the black hole fraction gradually increased. This inflated the cluster, enhancing tidal stripping and tail formation. A gigayear from now, the cluster will dissolve as a 100\% black hole cluster. Initially denser clusters end up with lower black hole fractions, smaller sizes, and no observable tails. Black hole-dominated, extended star clusters are therefore the likely progenitors of the recently discovered thin stellar streams in the Galactic halo.\\
\end{abstract}

%STARTmain
%STARTintro
In recent years, a few dozen thin ($\lesssim100$ pc) stellar tidal streams have been discovered in the Milky Way halo\cite{2016MNRAS.463.1759B,2018ApJ...862..114S,2018MNRAS.481.3442M,2019ApJ...872..152I}. Their elemental abundances and distribution in the Galaxy provide important constraints on the formation of the Galaxy and its dark matter distribution\cite{2010ApJ...712..260K}. Their narrow widths  imply that their progenitor stellar systems  had a low velocity dispersion and were dark matter-free star clusters rather than dark matter-dominated dwarf galaxies. However, a progenitor has not been found for any of these streams and the star cluster nature is questioned by two recent findings: firstly, the inferred mass-loss rate of the GD-1 stream is several times higher\cite{2020MNRAS.494.5315D} than what is found in frequently cited models of cluster evolution\cite{2003MNRAS.340..227B} and secondly, only mild tidal distortions and no tidal tidal tails were found\cite{2018MNRAS.473.2881K} for several globular clusters (GCs) with extremely radial orbits\cite{2018ApJ...863L..28M} passing through the strong tidal field near the Galactic center where tidal stripping is efficient. These results raise the questions of what drove the escape rate of the streams’ progenitors and why this mechanism is not active in all star clusters.

The metal-poor GC Palomar~5 (hereafter Pal~5) is one of the few known star clusters with extended tidal tails  associated with it\cite{2001ApJ...548L.165O},  spanning $\gtrsim20~$degrees on the sky, making it a Rosetta stone for understanding tidal tail/stream formation. The cluster has an unusually large half-light radius of $\sim20$~pc\cite{2017ApJ...842..120I} and combined with its relatively low mass of $\sim10^4~M_\odot$, its average  density  is among the lowest of all Milky Way GCs:  $\sim0.1~M_\odot/{\rm pc}^3$,  comparable to the stellar density in the solar neighbourhood. A  low density facilitates  tidal stripping and the formation of tidal tails\cite{2004AJ....127.2753D}, but it is not known whether this low density is the result of nature, or nurture.

It has been proposed that Pal 5 simply formed with a low density\cite{2004AJ....127.2753D} and has always been a collisionless system, meaning that two-body interactions were not important in its evolution. 
However, some properties of Pal 5 are reminiscent of other GCs, such as a spread in sodium abundances\cite{2002AJ....123.1502S} and a flat stellar mass function\cite{2017ApJ...842..120I}. These features have been attributed to high initial densities\cite{2017ApJ...836...80E,2018MNRAS.478.2461G} and collisional evolution\cite{2003MNRAS.340..227B} and suggest that Pal 5 in fact is, or was, a collisional stellar system, like the rest of the Milky Way GCs\cite{2011MNRAS.413.2509G}. In this study we aim to reconcile the  low density of Pal 5 with collisional evolution.

Since the discovery of gravitational waves\cite{2016PhRvL.116f1102A}, updated metallicity-dependent stellar wind and supernova prescriptions have been implemented in GC models\cite{2016PhRvD..93h4029R,2020PhRvD.102l3016A}. In these models, a large fraction of black holes (BHs) that form from massive stars have masses above $20~M_\odot$ and do not receive a natal kick, as the result of fallback of material,  damping the momentum kick resulting from asymmetries in the supernova explosion \cite{2001ApJ...554..548F,2012ApJ...749...91F}. The presence of a BH population in a star cluster accelerates its relaxation driven expansion\cite{2004ApJ...608L..25M,2008MNRAS.386...65M} and  escape rate\cite{2019MNRAS.487.2412G,2020MNRAS.491.2413W}. Observational motivation for considering the effect of BHs on GC evolution stems from the discovery of accreting BH candidates in several GCs with deep radio observations\cite{2012Natur.490...71S,2013ApJ...777...69C} and a BH candidate in a detached binary in NGC 3201\cite{2018MNRAS.475L..15G}. Here we investigate the possibility that Pal 5 was much denser in the past and that the present-day structure and prominent tidal tails are the result of a BH population.
%ENDintro

\section*{Results}
We perform star-by-star, gravitational $N$-body simulations with {\sc nbody6++gpu}\cite{2015MNRAS.450.4070W}. All clusters are evolved for 11.5 Gyr on the orbit of Pal 5 in a three-component Milky Way (bulge, disc, halo) and the simulations include the effect of stellar and binary evolution. No primordial binaries were included.
We consider two prescriptions for BH natal kicks. First we consider the most up-to-date BH recipes\cite{2020A&A...639A..41B}, in which  approximately 73\% of the  mass of the BH population is retained  after natal kicks, almost independently of the initial escape velocity of the cluster (see Methods). Then we test the collisionless hypothesis  and draw BH kick velocities from a Maxwellian with the same dispersion as that of  neutron stars (NSs). Because we need lower densities for these models, all BHs are ejected by the natal kicks in almost all models (in four models a single BH was retained). The latter approach is similar to that of Dehnen et al.\cite{2004AJ....127.2753D}, with the added effect of stellar evolution and a direct summation code to correctly include the effect of two-body interactions. We will  refer to these two sets of models as wBH and noBH, respectively. The initial parameters and all results are summarised in Tables 1 and 2.
 
We vary the initial number of stars ($N_0$) and the initial mass density within the half-mass radius: $\rho_{\rm h0}\equiv 3M_0/(8\pi r_{\rm h0}^3)$, where $M_0\simeq0.64~M_\odot\times N_0$ is the initial mass and  $r_{\rm h0}$ is the initial half-mass radius. We search for a model that best reproduces the observed number of  stars of Pal 5, $N_{\rm cluster}=1,550$,  and its half-light radius, $R_{\rm eff}=3.21~$arcmin/$18.7~$pc  (see Methods). $R_{\rm eff}$  is defined as the distance to the cluster center containing half the number of observed stars.  We first run a coarse grid of models, followed by a finer grid close to the parameters that give the best match in the coarse grid.  The best wBH model (wBH-1) has $N_0=2.1\times10^5$ and $\rho_{\rm h0}=80~M_\odot/{\rm pc}^3$. This cluster lost 92\% of its initial mass of $1.34\times10^5~M_\odot$ by stellar evolution and escapers and the density decreased nearly three orders of magnitude because of stellar evolution and dynamical heating by BHs. We note that the half-mass radius $r_{\rm h}\simeq18.8~$pc, as determined from the 3-dimensional mass distribution, is  similar to $R_{\rm eff}=18.2~$pc, which is determined from the projected distribution of observable stars. If mass follows light,  $R_{\rm eff} \simeq0.75r_{\rm h}$ and for mass segregated clusters without BHs it can be as small as $0.5r_{\rm h}$\cite{2007MNRAS.379...93H}. The fact that  $R_{\rm eff} \simeq r_{\rm h}$ for wBH-1 implies that the observable stars are less concentrated  than the mass profile. This is the result of the BH population which sinks to the center via dynamical friction against the lower mass stars\cite{2004ApJ...608L..25M,2008MNRAS.386...65M,2016MNRAS.462.2333P}, where they remain in a quasi-equilibrium distribution\cite{2013MNRAS.432.2779B}. The  surface density profile and the properties of the stream are in good agreement with the observations (Figure 1).
The small-scale density variations in the tails are not reproduced by our model (see Supplementary Figure 1). This is because they are likely the result of interactions with dark matter subhalos\cite{2017MNRAS.470...60E} or the Galactic bar and giant molecular clouds in the disc\cite{2019MNRAS.484.2009B}, which are not included in our model.
The observed line-of-sight velocity dispersion of stars in the tails is $2.1\pm0.4$~km/s\cite{2015MNRAS.446.3297K}, which is well reproduced by wBH-1  ($2.4\pm0.1$~km/s, for giants in the same region as the observations). The most striking property of wBH-1 is its large BH fraction at present: $f_{\rm BH}= 22\%$. We define $f_{\rm BH}$ as the total  mass in BHs that are bound to the cluster over the total bound cluster mass, that is, $f_{\rm BH}=M_{\rm BH}/M$. The BH population is made up of 124 BHs with an average mass of $17.2~M_\odot$ (that is, $M_{\rm BH} = 2,178~M_\odot$),  currently residing  within $R_{\rm eff}$ (see Figure~1). This $f_{\rm BH}$ is more than twice as large as what is expected from a canonical stellar initial mass function (IMF) and stellar evolution alone, and is the result of the efficient loss of stars over the tidal boundary, while the BHs were mostly retained because they are in the center. 

In all wBH models, $f_{\rm BH}$  increases in the first $\sim10$ Myr because of BH formation and stellar evolution mass loss. Then, $f_{\rm BH}$ decreases in the following $\sim100$ Myr because of BH ejections from the core\cite{2013MNRAS.432.2779B}. This happens because binary BHs form and interact with other BHs when two-body relaxation becomes important. These binaries become more bound in these interactions, eventually ejecting BHs and themselves from the cluster. What happens next depends on $\rho_{\rm h0}$: in dense clusters, most BHs are ejected before the cluster dissolves and these GCs have escape rates and  $R_{\rm eff}$ comparable to what is found in models of clusters without BHs (Figure 2). Clusters with lower initial densities have larger relaxation times, resulting in fewer BH ejections, while tidal stripping of stars is more efficient, leading to an increasing $f_{\rm BH}$ until $100\%$\cite{2011ApJ...741L..12B} (Figure 2). For $f_{\rm BH}\gtrsim0.1$,  $R_{\rm eff}\sim15$~pc, comparable to what is found for about half of the GCs in the outer halo of the Milky Way\cite{2010MNRAS.401.1832B}, suggesting that these `fluffy' GCs are BH rich and candidates to produce prominent stellar streams. This idea is further supported by the strong correlation between $f_{\rm BH}$ and $R_{\rm eff}$ and the fraction of stars in the stream (Figures 3 and 4). 
Theory suggests\cite{2013MNRAS.432.2779B} that in idealised single-mass star clusters that fill their tidal radius there exists a critical $f_{\rm BH}\simeq10\%$ at which the mass-loss rate of stars and BHs is the same and $f_{\rm BH}$ remains constant at $10\%$ while the cluster loses mass. For higher(lower) $f_{\rm BH}$, stellar mass is lost at a higher(lower) rate by tidal striping than BH mass is lost by ejections from the core. This implies that clusters can evolve to 100\% BH clusters if they  form with $f_{\rm BH}\gtrsim10\%$ and can remain above it during their evolution. Our  models suggest that for multimass models this critical fraction is lower: $\sim2.5\%$. Since  for a canonical IMF the initial $f_{\rm BH}\simeq5-10\%$, depending on metallicity, it is possible for some GCs to remain above the critical $f_{\rm BH}$ and evolve to 100\% BH clusters\cite{2011ApJ...741L..12B}, as we find in our wBH-1 model. Because $f_{\rm BH}$ always evolves away from 10\%,  the distribution of $f_{\rm BH}$ values of a GC population becomes bimodal. Whether cluster evolve towards BH-free clusters, or 100\% BH clusters depends on their initial density relative to the tidal density. Because $f_{\rm BH}$ affects the density, a unimodal initial density distribution can evolve towards a bimodal preseny-day density distribution, as is observed in the halo\cite{2010MNRAS.401.1832B}. 

We now discuss the  noBH models. The best-fit  model (noBH-1) has $N_0=3.5\times10^5$ and $\rho_{\rm h0}=9.625~M_\odot/{\rm pc}^3$, that is, approximately twice as massive and an order of magnitude less dense than wBH-1. The resulting observable parameters $N_{\rm cluster}(R_{\rm eff})$ are within $27\%(5\%)$ of the values of Pal 5. The resulting cluster density profile and stream properties are similar to those of wBH-1 (Supplementary Figure 1), hence based on these observables it is not possible to prefer either of the assumptions for  BH kicks. The wBH-1 model predicts a higher central velocity dispersion of 580~m/s vs. 350~m/s for noBH-1 (see Methods). The inferred central dispersion from the literature compilation by Baumgardt \& Hilker\cite{2018MNRAS.478.1520B} is $550^{+150}_{-110}~$m/s; that is, favouring the BH hypothesis.
We quantified the degree of fine-tuning of $N_0$ and $\rho_{\rm h0}$ that is required to obtain the best model in both cases. For the wBH models, we find that the uncertainty in the present-day properties can be covered by a relatively large range of initial densities, while for the noBH models we find that variations in the initial density are amplified by more than an order of magnitude in variations in the present-day properties (see Supplementary Information). This means that a relatively large range of initial conditions in wBH models lead to similar present-day properties, while  for the noBH models a high degree of fine-tuning in the initial density is required to obtain noBH-1. In addition, noBH-1 completely dissolves at 11.8 Gyr, that is, 300 Myr after we observe the cluster, while wBH-1 survives for another 1.4 Gyr. By assuming simple power-law distributions for the initial cluster masses and densities, we estimate that the probability of finding a cluster with the properties of Pal 5 for wBH-1(noBH-1) is $1/200(1/3.6\times10^5)$. Given that the Milky Way has $\sim150$ GCs, the wBH-1 model provides the more likely explanation for Pal 5. The velocity dispersion and the fine-tuning and timing arguments all favor the BH hypothesis. The final argument in support of the BH hypothesis is that the higher initial density and the resulting collisional nature of Pal 5 make it easier to understand the flat stellar mass function, its multiple populations and its relation to the rest of the Milky Way GC population. 

In the observed mass range ($0.6-0.8~M_\odot$), the mass function of Pal 5 is flatter ($dN/dm\propto m^{\alpha}$, with $\alpha\simeq-0.5$\cite{2017ApJ...842..120I}) than what is expected from a canonical IMF ($\alpha=-2.3$), suggesting that Pal 5 has  preferentially lost low-mass stars. We note that this result has been questioned and the mass function slope may actually be close to the initial value (Iskren Georgiev, private communication). In Figure 5 we show the mass function of bound stars and remnants of wBH-1. For main sequence stars $>0.5~M_\odot$ it has a slope of $\alpha=-1.5^{+0.2}_{-0.4}$, which is flatter than the IMF, but steeper than the observed slope. The noBH-1 model has a  slope comparable to wBH-1, implying that in the later parts of its evolution the cluster became collisional. The similarity in mass function slope between wBH-1 and noBH-1 means that we can not use the observed mass function to distinguish between the two scenarios. However, in the noBH models the only way to reconcile the models with the observations is to start with a flatter IMF, because the alternative (that is, reducing the initial relaxation timescale by increasing the initial density) leads to a $R_{\rm eff}$ that is too small.  The BH scenario, however,  leaves the possibility to increase the initial cluster density and still end up with the same present-day BH fraction, $R_{\rm eff}$ and tidal tails. 
For example, tidal heating by interstellar gas clouds in the early  evolution is probably important in the evolution of GCs\cite{2010ApJ...712L.184E,2015MNRAS.454.1658K}, but this effect is not included in our models.
This tidal heating leads to a decrease in the cluster mass\cite{1958ApJ...127...17S,2006MNRAS.371..793G} and for clusters with a low central concentration the cluster density also decreases\cite{2016MNRAS.463L.103G}. BHs sink to the cluster center by dynamical friction on a fraction of a relaxation time and if this is shorter than the tidal heating timescale, then the BHs are less affected by this mechanism of tidal stripping than the stars, such that $f_{\rm BH}$ increases, counteracting the reduction of $f_{\rm BH}$ from BH ejections from the core due to the higher density. This would be a pathway to start with higher densities. Alternatively, something may have happened to Pal 5 at a later stage of its evolution. It is likely that Pal~5 formed in a dwarf galaxy that was accreted onto the Milky Way, which is supported by its tentative association with the Helmi streams\cite{2019A&A...630L...4M}. The cluster could have lost a fraction of its  loosely bound stars as a result of the removal of its host galaxy\cite{2015MNRAS.447L..40B}, thereby increasing  $f_{\rm BH}$.
Finally, a flatter IMF  at high masses would also lead to a higher $f_{\rm BH}$. This also affects the evolution, because a flatter IMF would lead to faster expansion and a higher mass loss rate\cite{2017ApJ...834...68C, 2019MNRAS.487.2412G, 2020MNRAS.491.2413W}.
 In all these scenarios, denser initial conditions and therefore more equipartition among low-mass stars, whilst ending with the same present-day $f_{\rm BH}$ is a possibility. Understanding the interplay between early mass loss, mass segregation, the (high-mass) IMF and accretion on the Milky Way to find limits on the allowed initial density is an interesting topic for a follow-up study. The initial density is a critical ingredient for our understanding of GC formation\cite{2017ApJ...836...80E,2018MNRAS.474.4232K,2018MNRAS.478.2461G}, evolution\cite{2015MNRAS.454.1658K}, BH natal kicks\cite{2012ApJ...749...91F} and binary BH mergers\cite{2016PhRvD..93h4029R, 2020ApJS..247...48K,2020PhRvD.102l3016A}.

Our results have implications for our understanding of GC evolution. In frequently cited models of cluster evolution without BHs, the relaxation driven escape rate on the orbit of Pal~5 is about $\sim1~M_\odot/{\rm Myr}$ (Figure 2). We note that our noBH models have much higher escape rates, but this is because they start with lower initial densities than what is usually done\cite{2003MNRAS.340..227B}. It is well-established that the mass loss rate in these models is insufficient to `turn-over' a power-law initial cluster mass function with index $-2$, as is observed for young star clusters in the nearby Universe, into the observed peaked (logarithmic) mass distribution with a typical mass of $2\times10^5~M_\odot$: too many low-mass GCs survive beyond $\gtrsim 5~$kpc\cite{2001MNRAS.322..247V}. 
The required escape rate is about an order magnitude larger: $20~M_\odot/{\rm Myr}$\cite{2001ApJ...561..751F}. From our models we can conclude that Pal~5 is currently losing mass at that rate  (Figure 2), hence if a large fraction of GCs go through a similar evolutionary phase, then relaxation driven evaporation is more important in shaping the GC mass function than usually assumed, reducing the need for additional GC disruption mechanisms\cite{2010ApJ...712L.184E, 2015MNRAS.454.1658K} or a peaked initial cluster mass function. Because variations in $f_{\rm BH}$ lead to an order of magnitude variation in the escape rate (Figure 2), we conclude that the effect of BHs is of comparable importance in shaping the GC mass function as the details of the Galactic orbit.

\section*{Discussion}
We now consider our results in the context of the  Milky Way GC population. About  $10\%$  of the Milky Way GCs have  $R_{\rm eff}\gtrsim10~$pc, which led some authors to suggest that two-body relaxation is not important and that their evolution is collisionless\cite{2004AJ....127.2753D,2011ApJ...726...47S}. These fluffy GCs are found predominantly at large Galactocentric radii and have low masses ($\lesssim10^5~M_\odot$). More specifically, beyond $\gtrsim8~$kpc from the Galactic center, half of the GCs  have $R_{\rm eff}=10-20$ pc\cite{2010MNRAS.401.1832B}.  Our results suggest that these large and low-density GCs are  BH rich. Their low densities, relative to the tidal densities on their orbits can be used to  constrain their initial densities and masses in a similar way as we did for Pal 5. In wBH-1, about half of the observable stars are lost in the last 3 Gyr and in this period the stream became visible above the background. Combined with the remaining lifetime of $\sim1$~Gyr, Pal~5 has an observable stream for 30\% of its lifetime.
There are $26$ low-density GCs beyond 8~kpc from the Galactic center and for about half of them tidal tails or tidal features have been found (Supplementary Table 1). Of these, there are 12 that are at distances comparable to Pal 5, or closer. If we assume that the escape rate is the same for these GCs, combined with the fact the GC mass function is uniform at low masses, then about 30\% of these GCs are in the final 30\% of their evolution. From this we expect that $\sim4$ GCs to have prominent tidal tails, providing an explanation for the rarity of Pal 5 and the low-number of known GCs with streams. We can now make an order of magnitude estimate of how many streams the Milky Way fluffy GC population has generated in the past. The $\sim10$ fluffy, nearby, GCs all have masses below $10^5~M_\odot$, hence the  mass function of fluffy clusters is $dN/dM \simeq 10^{-4}~M_\odot^{-1}$. If these GCs all lose mass at a constant rate of $20~M_\odot/{\rm Myr}$, then this mass function is constant in time and we expect that these GCs  contributed $2\times10^{-3}$ streams per Myr, or about 20 streams in the last 10 Gyr. This estimate supports the idea that the $\sim30$\cite{2016MNRAS.463.1759B,2018ApJ...862..114S,2018MNRAS.481.3442M,2019ApJ...872..152I} known cold streams in the  halo resulted from BH-rich, extended GCs that dissolved in the Milky Way halo. 
 
We conclude with a discussion on  observational tests of the BH hypothesis. We computed the   rate of microlensing events of background quasars in the field of Pal 5 and find that the event rate is too low ($\sim10^{-8}/$yr).  We then looked at the effect of BHs on stellar kinematics and found that it is possible to infer the BHs from the kinematics of the stars. For a BH fraction of 22\%, the velocity dispersion of giant stars within $R_{\rm eff}$ is $580~$m/s, which is $\sim200$~m/s higher than for a cluster without BHs (Figure 6).  
The significance of the available velocity dispersion measurement of $550^{+150}_{-110}~$m/s\cite{2018MNRAS.478.1520B} can be improved by increasing the sample of stars, and constraining the properties of binary stars, including their orbital periods. 
We estimate an upper limit for the periods of binaries with giant stars of  $P\simeq0.5~$yr, because longer period binaries were ionised by interactions with BHs. Finding a binary with a larger period would challenge our BH hypothesis, while the absence of such binaries could in combination with additional $N$-body model with primordial binaries be used as support for it. Finally, BHs with stellar binary companions may also exist\cite{2018MNRAS.475L..15G} and could be found from  their large velocity variations.  A multi-epoch observing campaign to obtain line-of-sight velocities  is  therefore needed to establish  with high precision the central velocity dispersion and the properties of the binaries (see Methods for details). This would provide the critical test of the hypothesis that Pal 5 hosts a $\gtrsim2\times10^3~M_\odot$  population of stellar-mass BHs.
%ENDmain

\section*{Data availability.}
A snapshot of the wBH-1 model is published on zenodo (\href{https://zenodo.org/record/4739181\#.YJK0Q0gzZJw}{doi:10.5281/zenodo.4739181}). All $N$-body data are available upon request.

\section*{Code availability.}
\noindent {\sc nbody6++gpu} is available from \href{https://github.com/nbodyx/Nbody6ppGPU}{https://github.com/nbodyx/Nbody6ppGPU}. {\sc limepy} is available from \href{https://github.com/mgieles/limepy}{https://github.com/mgieles/limepy}.

%%% figures
\newpage\begin{figure}
\centering
\includegraphics[width=16cm]{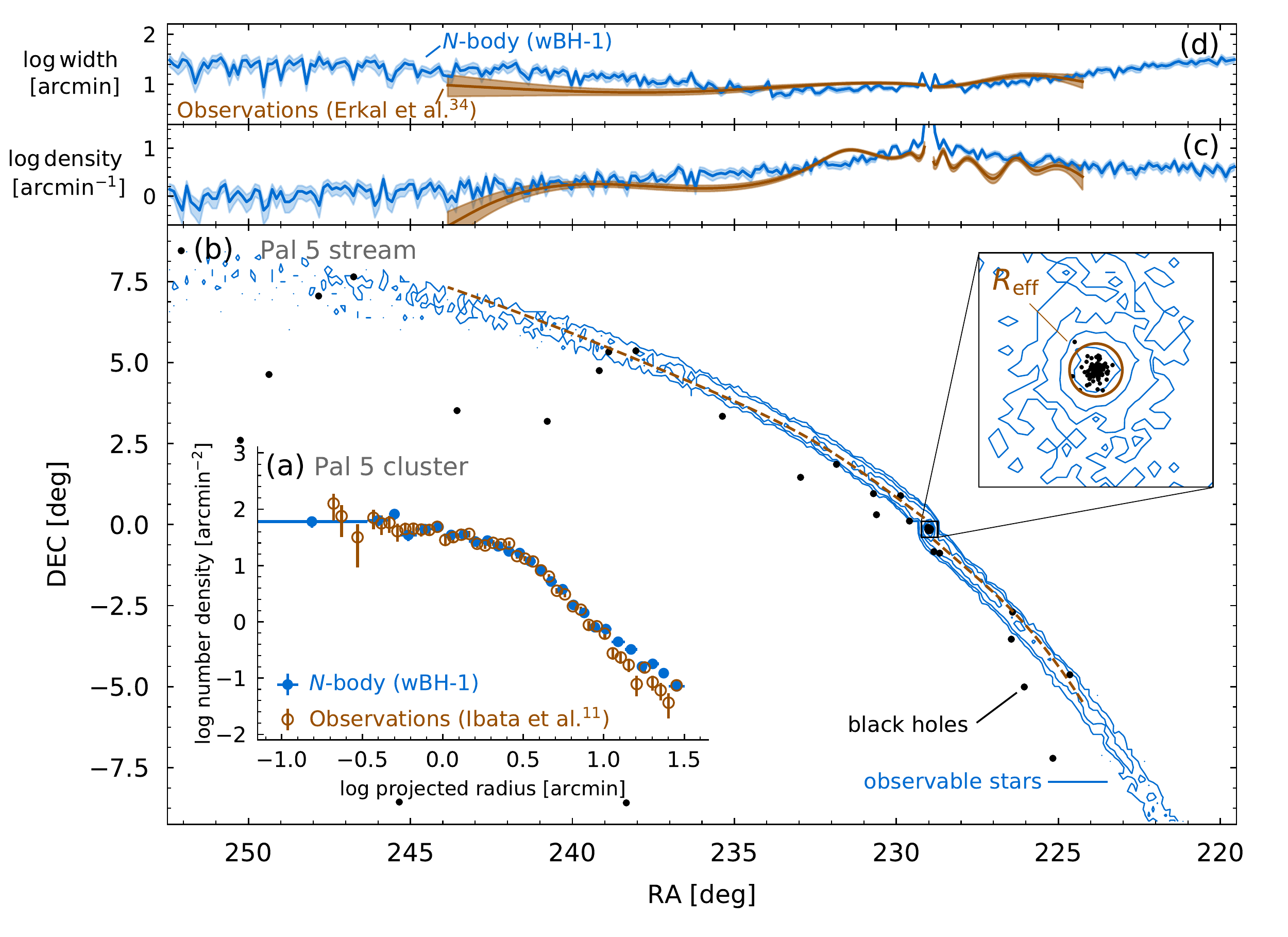}
\caption{{\bf Comparison between the $N$-body model wBH-1 and observations of Pal~5 and its stream.} (a) Density profile of observable stars in wBH-1, which provides an excellent match to the observations. There is also good agreement between the stellar stream track (b), the stream density and its gradient (c) and stream width (d) between observations and wBH-1, implying that the rate of escape of stars from Pal 5 and their velocity dispersion in the last few Gyrs are correctly reproduced by wBH-1. The blowout of the cluster shows that almost all of the 124 bound BHs  are within $R_{\rm eff}$. All error bars denote the 67\% confidence interval.}
\end{figure}

\begin{figure}
\centering\includegraphics[width=12cm]{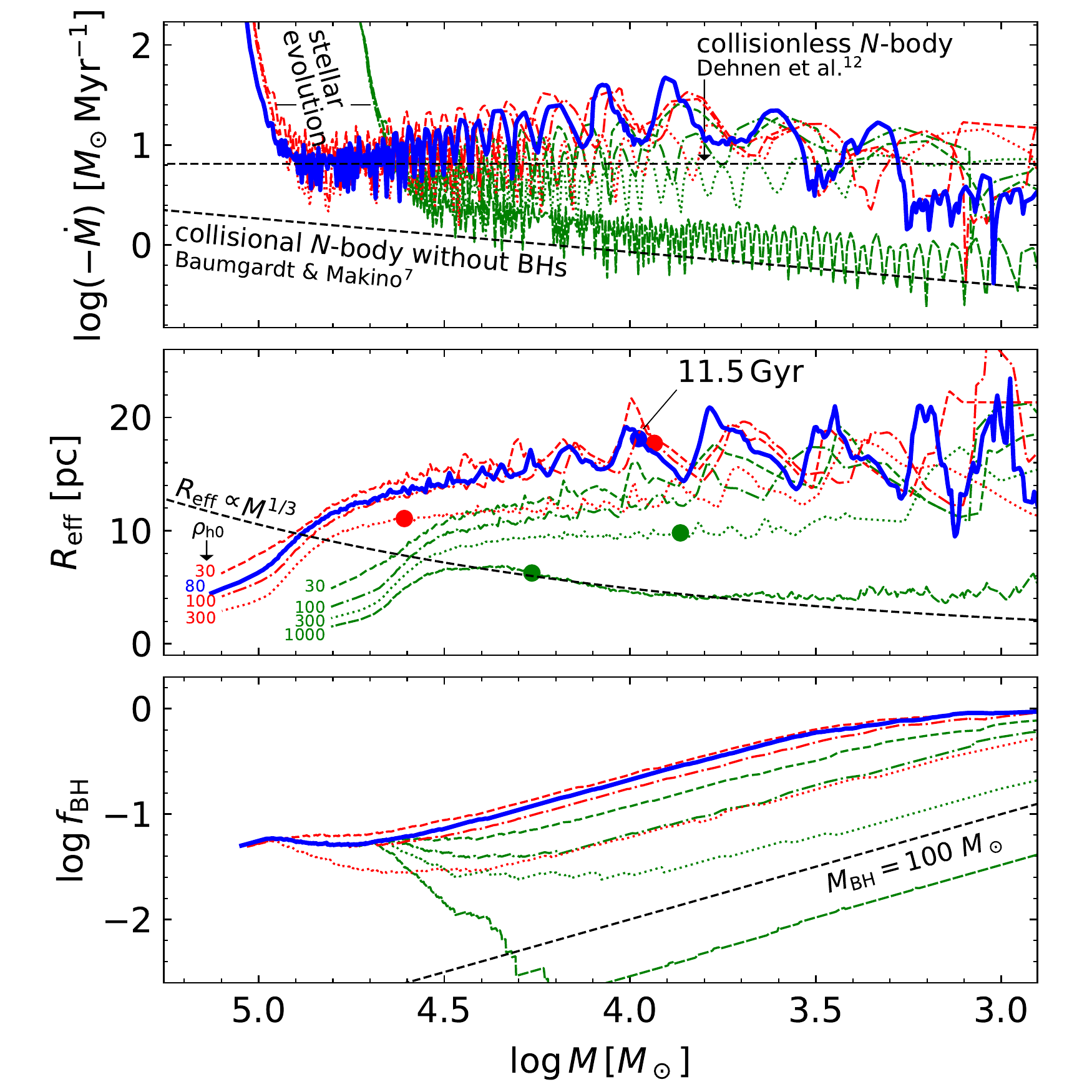}
\caption{{\bf Results for different initial conditions for clusters with BHs. } Results of wBH-1 (blue, thick lines), compared to other wBH models with $N_0=100k$ (green lines) and $200k$ (red lines) with different initial densities. Different panels show the variation of diagnostic quantities as a function of  bound cluster mass ($x$-axis), which decreases in time. Top: Total (positive)  mass-loss rate ($-\dot{M}$), which is initially high as the result of stellar evolution. After about 30\% of the mass is lost, $\dot{M}$ is dominated by escaping stars and BHs. The results of the collisionless $N$-body model of Pal~5\cite{2004AJ....127.2753D} and that of the  collisional $N$-body models without BHs\cite{2003MNRAS.340..227B} are overplotted. Because of the BHs, wBH-1 loses mass at a similar rate as the lower density collisionless models. Middle: Evolution of $R_{\rm eff}$ with the age of 11.5 Gyr indicated with dots. The models do not evolve towards a constant luminosity density (that is, $R_{\rm eff}\propto M^{1/3}$), as predicted for single-mass clusters\cite{1961AnAp...24..369H}.
Instead, $R_{\rm eff}$ remains approximately constant while the cluster evolves to lower mass and  there is a factor of five spread in $R_{\rm eff}$ at a given mass due to variations in $f_{\rm BH}$ (bottom). Dense clusters eject all BHs, while low-density clusters lose stellar mass at a higher rate than BH mass, and evolve along tracks of nearly constant BH mass towards a 100\% BH cluster. }
\end{figure}

\begin{figure}
\centering\includegraphics[width=12cm]{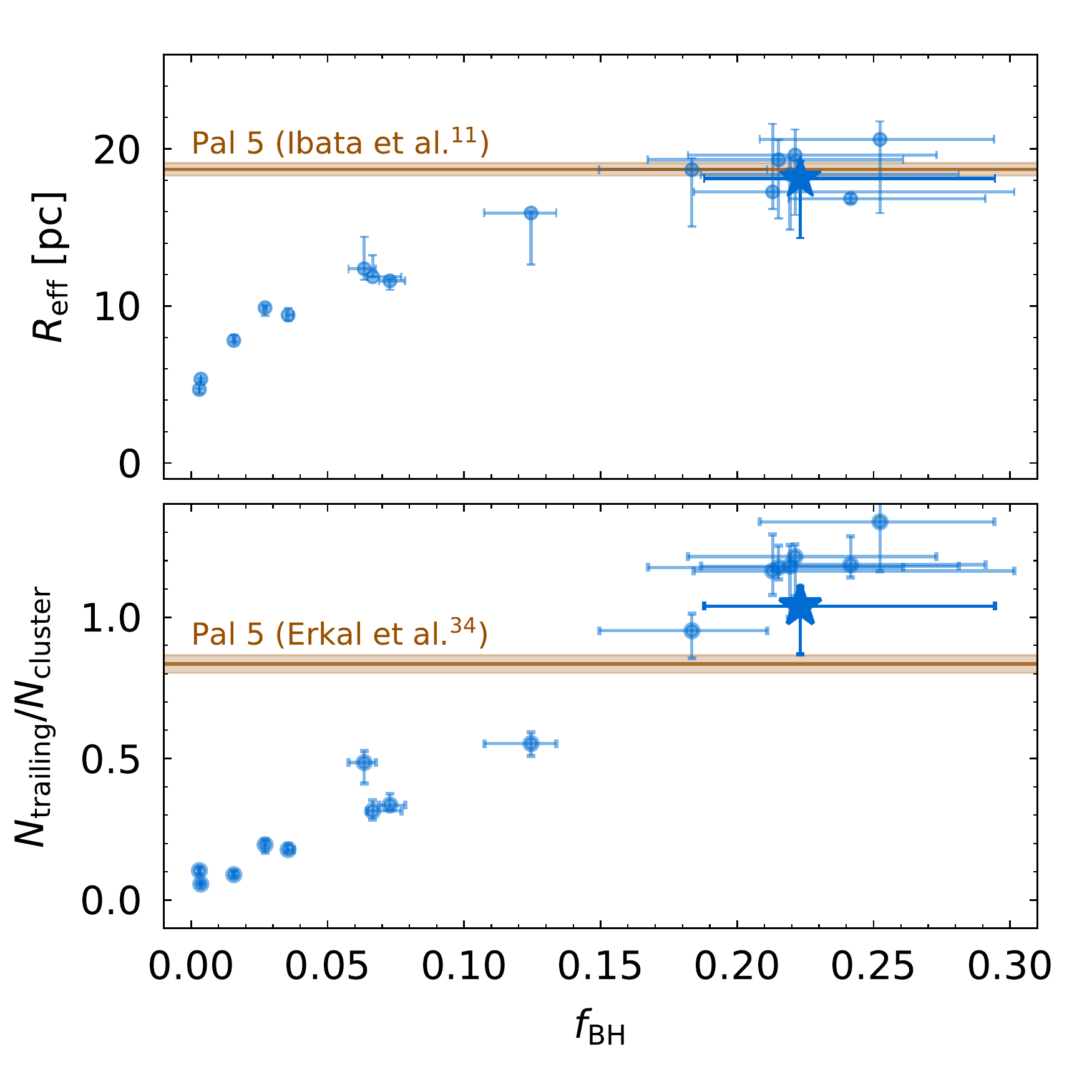}
\caption{{\bf Dependence of cluster and stream properties on the BH content.} Effective radius (top) and the relative number of stars in the trailing tail (bottom) for all surviving wBH clusters when their mass is $0.95\times10^4~M_\odot$ (the mass of wBH-1 at 11.5 Gyr) as a function of $f_{\rm BH}$ (blue dots with error bars). Observed values\cite{2017ApJ...842..120I, 2017MNRAS.470...60E} are shown as horizontal lines with the shaded region indicating the 1$\sigma$ confidence interval. Values of wBH-1 are shown by large stars. Here $N_{\rm trailing}$ is defined as the number of observable stars in the trailing tail, within 4$^{\circ}$ of right ascension of Pal 5, which in the model corresponds to stars that escaped in the preceding $\sim1$ Gyr. Because these quantities vary along the orbit at these low masses, we plot  the range within the nearest peri and apo as error bars. 
The strong correlation of both quantities with $f_{\rm BH}$ suggests that for a given orbit and cluster mass, the BH content sets both $R_{\rm eff}$ of the cluster and the prominence of  the tidal tails.}
\end{figure}

\begin{figure}
\centering\includegraphics[width=12cm]{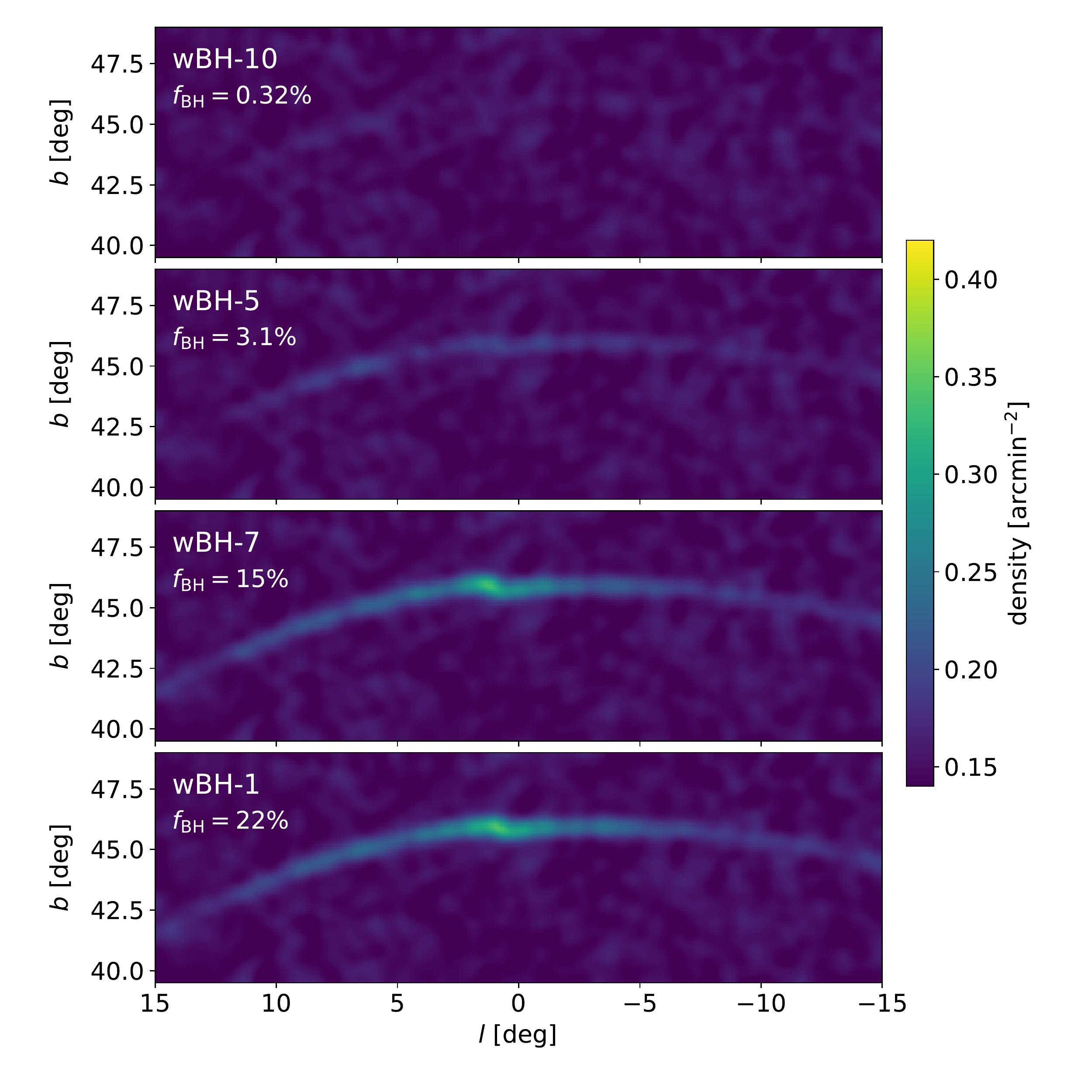}
\caption{{\bf Dependence on stream visibility on BH content.} Density map in  Galactic latitude ($l$) and longitude ($b$) of unbound stars of four different wBH models (indicated in the top left corners) with different $f_{\rm BH}$ at 11.5 Gyr. A density of background stars of $0.15/{\rm arcmin}^2$ was added to each model. Models with $f_{\rm BH}\lesssim10\%$ do not have prominent tidal tails.}
\end{figure}

\begin{figure}
\centering\includegraphics[width=12cm]{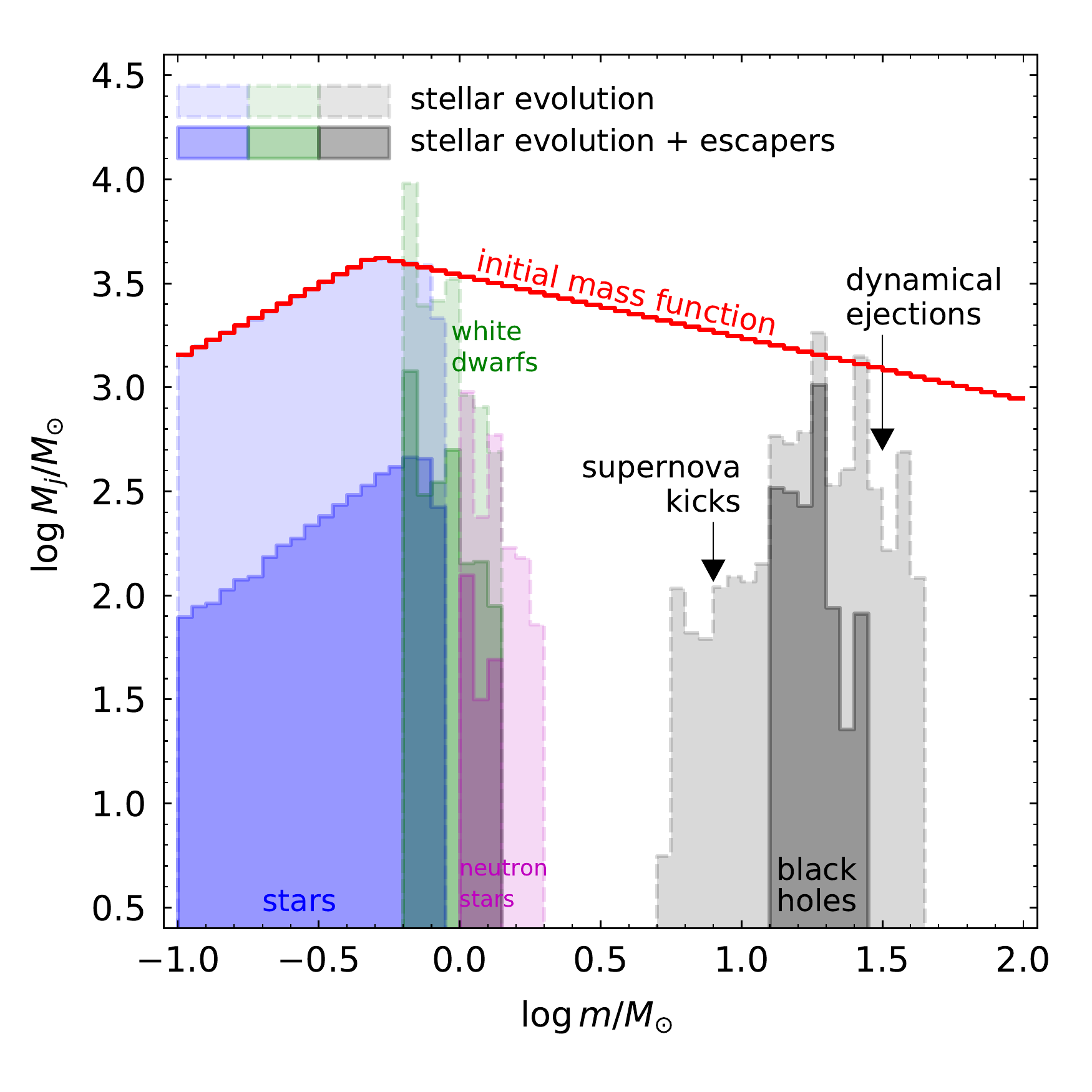}
\caption{{\bf Mass function of stars and remnants.} Mass function of bound stars and remnants of wBH-1 at 11.5 Gyr. The mass in each bin ($M_j$) is plotted for the IMF,  the mass function after only stellar evolution has been applied (light shading) and for the final results of the $N$-body simulation  (dark shading).  There is a  flattening of the stellar mass function due to preferential escape of low-mass stars.}
\end{figure}

\begin{figure}
\centering\includegraphics[width=15cm]{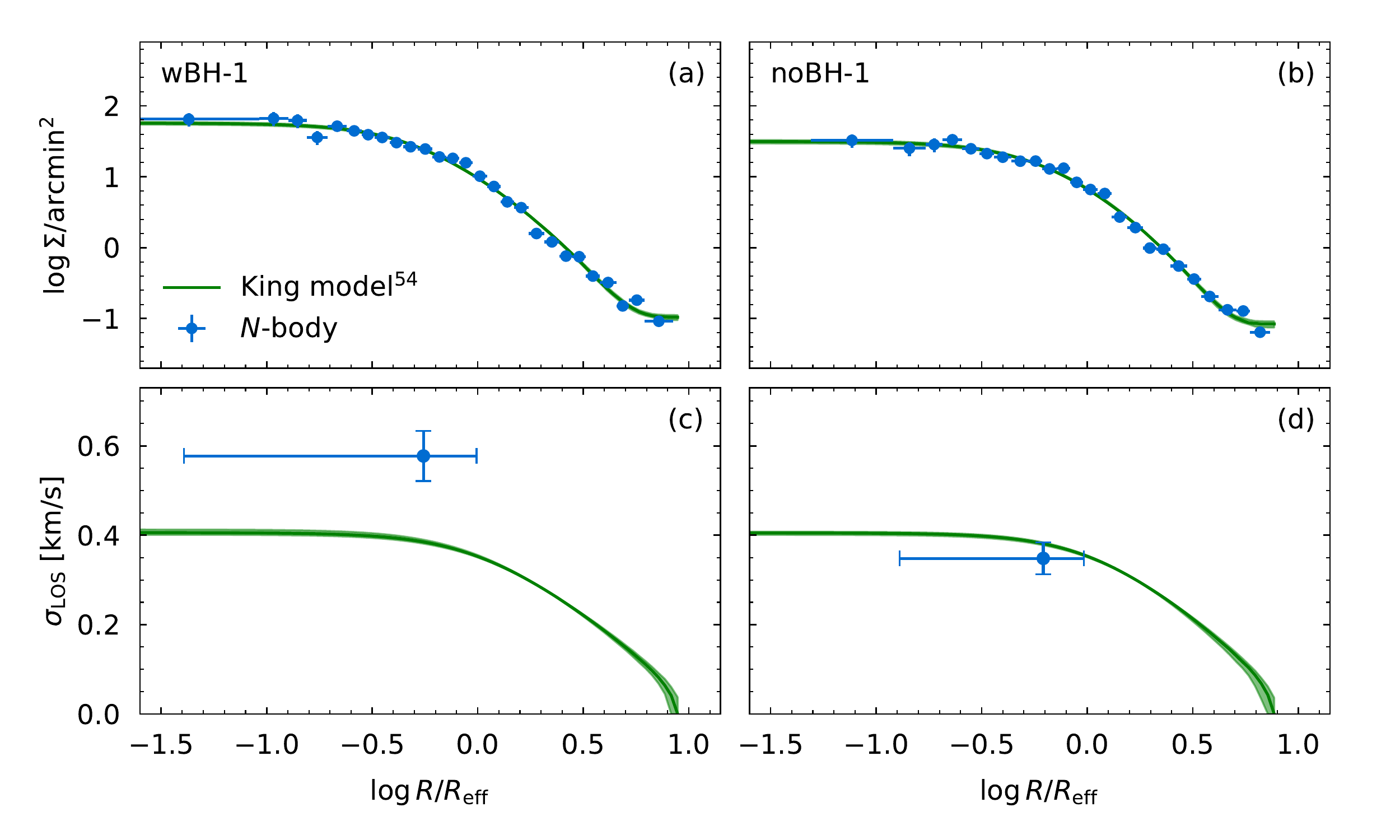}
\caption{{\bf Surface density  and velocity dispersion profiles of models with and without BHs.} Fits of King models\cite{1966AJ.....71...64K} plus a constant background (green lines and shaded region representing the $1\sigma$ uncertainty) to the surface density profile of observable stars (blue circles with error bars indicating the 67\% confidence interval) for wBH-1 ({\bf a}) and noBH-1 ({\bf b}).  ({\bf c,d},) The resulting velocity dispersion ($\sigma_{\rm LOS}$) of the model fits scaled to a total mass of $5\times10^3~M_\odot$ and half-mass radius of $25~$pc  assuming that the total mass is due to all of the stars, white dwarfs and neutron stars; that is, no BHs. Blue points with error bars in {\bf c} and {\bf d} are the dispersions in the $N$-body models of the giants within $R_{\rm eff}$ scaled to the same mass and half-mass radius for wBH-1 and noBH-1, respectively. For noBH-1 the dispersion of the giants is $0.35\pm0.04$, consistent with the King model prediction. For wBH-1, this dispersion is $0.58\pm0.06$~km/s, or 3$\sigma$ above the central dispersion prediction from the King model.  With precise line-of-sight velocities and a photometric mass estimate, this difference is detectable.}
\end{figure}

\def\aap{{\it Astron. Astrophys.}}
\def\pasj{\it Publ. Astron. Soc. Jpn}
\def\pasp{\it Publ. Astron. Soc. Pac.}
\newcommand{\apj}{Astrophys. J.}
\newcommand{\apjs}{Astrophys. J. Suppl.}
\newcommand{\apjl}{Astrophys. J. Letters}
\newcommand{\aj}{Astron. J.}
\newcommand{\araa}{Annu. Rev. Astron. Astrophys.}
\newcommand{\mnras}{{\it Mon. Not. R. Astron. Soc.}}
\newcommand{\nat}{Nat.}
\newcommand{\prd}{Phys. Rev. D}

\newpage

\begin{addendum}
 \item MG and EB acknowledge financial support from the European Research Council (ERC StG-335936, CLUSTERS) and MG acknowledges support from the Ministry of Science and Innovation through a Europa Excelencia grant (EUR2020-112157).
FA acknowledges support from a Rutherford fellowship (ST/P00492X/2) from the Science and Technology Facilities Council. EB acknowledges financial support from a Vici grant from the Netherlands Organisation for Scientific Research (NWO).  MG thanks Mr Gaby P\'{e}rez Forcadell for installing the GPU server at the ICCUB on which all the simulations were run. The authors thank Rodrigo Ibata for sharing the data of Pal 5's surface density profile, {\L}ukasz  Wyrzykowski for discussions on microlensing and  Sverre Aarseth, Keigo Nitadori and Long Wang for maintaining {\sc nbody6} and {\sc nbody6++gpu}  and making the codes publicly available. MG and FA thank Long Wang and Sambaran Banerjee for discussions on the recent {\sc sse} and {\sc bse} updates and the implementation in {\sc nbody6++gpu}. This research made use of {\sc astropy}, a community-developed core Python package for Astronomy\cite{astropy:2013, astropy:2018}, {http://www.astropy.org}. 

\item[Author contributions] M.G. ran all $N$-body simulations, analysed them and was in charge of the writing. D.E. was in charge of stream modelling and deriving the orbit of Pal 5 and the parameters of the MW model. F.A. contributed to the BH physics of the $N$-body models. E.B. converted stream models to observed quantities and J.P. contributed to the binary properties. All authors assisted in the development, analysis and writing of the paper. 

 \item[Competing Interests] The authors  have no competing financial interests.

\item[Correspondence] Reprints and permissions information
is available at www.nature.com/reprints. Correspondence and requests
for materials should be addressed to
MG (mgieles@icc.ub.edu).
\end{addendum}

%%
%% TABLES
%%
%% If there are any tables, put them here.
%%

\newpage
\begin{landscape}
\begin{spacing}{1.0}
\begin{center}
\small
\begin{longtable}{ c c c c c c c c c c c c c c} 
 \hline\hline
 (1) & (2) & (3) & (4) & (5) & (6) & (7) & (8)   & (9)  & (10)& (11)& (12)& (13) & (14)\\ 
  $N_0$ & $\rho_{\rm h0}$     & $M_0$& $r_{\rm h0}$ &  $T_{\rm 1,550}$& $N_{\rm cluster}$  & $R_{\rm eff}$  & $M$ & $r_{\rm h}$& $f_{\rm BH}$ &$\delta_N$ & $\delta_R$& $\delta$ &Model \\  
  {[$10^3$]} &[$M_\odot/{\rm pc}^{3}$]&[$10^5M_\odot$]&[pc]& [Gyr]&    & [pc]& [$10^4M_\odot$]  & [pc]&[\%] & [\%] &[\%]& [\%] &\\\hline 
     50 &     30 &      0.319 &  5.04 &  2.74 &    -- &     -- &    -- &    -- &    -- &   --&    -- &   -- & wBH-17 \\
     50 &    100 &      0.319 &  3.38 &  4.06 &    -- &     -- &    -- &    -- &    -- &   --&    -- &   -- & wBH-16 \\
     50 &    300 &      0.319 &  2.34 &  5.58 &    -- &     -- &    -- &    -- &    -- &   --&    -- &   -- & wBH-14 \\
    50 &   1000 &      0.319 &  1.57  &  8.34 &      1052 &    4.05 &  0.483 &   5.29 & 0.701 & -32.1 & -78.3 & 84.7 & wBH-9\\
    100 &     30 &      0.638 &  6.35 &  5.40 &    -- &     -- &    -- &    -- &    -- &   --&    -- &   -- & wBH-15 \\
    100 &    100 &      0.638 &  4.25 &  7.85 &    -- &     -- &    -- &    -- &    -- &   --&    -- &   -- & wBH-13 \\
   100 &    300 &      0.638 &  2.95  & 11.36 &      1448 &     9.8 &  0.732 &   12.8 & 3.09 & -6.58 & -47.6 & 48 & wBH-5\\
   100 &   1000 &      0.638 &  1.97  & 17.99 &      3541 &    6.26 &   1.84 &   9.04 & 0.321 & +128 & -66.5 & 145 & wBH-10\\
    200 &     30 &       1.28 &  7.98 &  8.12 &    -- &     -- &    -- &    -- &    -- &   --&    -- &   -- & wBH-12 \\
   200 &    100 &       1.28 &  5.34  & 11.43 &      1399 &    17.7 &   0.86 &     20 & 20.1 & -9.74 & -5.3 & 11.1 & wBH-3\\
   200 &    300 &       1.28 &   3.7  & 17.64 &      7210 &    11.1 &   4.05 &   14.9 & 2.75 & +365 & -40.8 & 367 & wBH-11\\
   210 &     80 &       1.34 &  5.85  & 11.47 &      1497 &    18.1 &  0.949 &   18.8 & 22.3 & -3.42 & -3.11 & 4.62 & wBH-1\\
   215 &     80 &       1.37 &   5.9  & 11.13 &       459 &    19.9 &  0.419 &   15.3 & 44.6 & -70.4 & +6.15 & 70.7 & wBH-8\\
   220 &     70 &        1.4 &  6.21  & 11.45 &      1411 &    18.5 &  0.953 &   19.8 & 21.9 & -8.92 & -0.948 & 8.97 & wBH-2\\
   220 &     80 &        1.4 &  5.94  & 11.31 &       791 &    20.8 &  0.604 &   19.4 & 33 & -49 & +11.5 & 50.3 & wBH-6\\
   225 &     70 &       1.44 &  6.26  & 11.29 &      1091 &    16.6 &   0.77 &   16.5 & 29.1 & -29.6 & -11.4 & 31.7 & wBH-4\\
   225 &     80 &       1.44 &  5.98  & 11.71 &      2372 &    16.8 &   1.45 &   19.8 & 15.1 & +53.1 & -10 & 54 & wBH-7\\
 \hline\hline
 \caption{Overview of the 17 wBH $N$-body simulations of Pal 5. The different  columns present: (1) initial number of stars; (2) initial half-mass density; (3) initial mass; (4) initial half-mass radius;  (5) age when the cluster had the observed number of stars ($N_{\rm cluster}=1,550$); (6)
observable number of stars at 11.5 Gyr; (7)  half-light radii at 11.5 Gyr; 
(8) bound mass at 11.5 Gyr; (9) half-mass radius of the bound stars at 11.5 Gyr; (10) BH mass fraction at an age of 11.5 Gyr; (11) fractional difference with $N_{\rm cluster}$; (12) fractional difference with observed $R_{\rm eff}=18.7\pm0.4$\,pc; (13) overall fractional difference $\delta=\sqrt{\delta_N^2 + \delta_R^2}$; (14) Model ID, sorted in level of agreement with the data with wBH-1 being the closest match.  Dashes are used for models that dissolve before 11.5\,Gyr.}
\end{longtable}
\end{center}
\end{spacing}

\begin{spacing}{1.0}
\begin{center}
\small
\begin{longtable}{ c c c c c c c c c c c c c c} 
 \hline\hline
 (1) & (2) & (3) & (4) & (5) & (6) & (7) & (8)   & (9)  & (10)& (11)& (12)& (13) & (14)\\ 
  $N_0$ & $\rho_{\rm h0}$     & $M_0$& $r_{\rm h0}$ &  $T_{\rm 1,550}$& $N_{\rm cluster}$  & $R_{\rm eff}$  & $M$ & $r_{\rm h}$& $f_{\rm BH}$ &$\delta_N$ & $\delta_R$& $\delta$ & ID\\  
  {[$10^3$]} & [$M_\odot/{\rm pc}^{3}$]&[$10^5M_\odot$]&[pc]& [Gyr]&    & [pc]& [$10^4M_\odot$]  & [pc]&[\%] & [\%] &[\%]& [\%] &\\\hline 
   100 &    10 &      0.638 &  9.15  &  7.84 &       609 &    4.42 &  0.289 &   5.76 & 0 & -60.7 & -76.4 & 97.5 & noBH-14\\
   100 &    30 &      0.638 &  6.35  & 15.09 &      2682 &    4.71 &   1.49 &   7.24 & 0 & +73.1 & -74.8 & 105 & noBH-15\\
   100 &   100 &      0.638 &  4.25  & 16.71 &      3607 &    3.57 &   2.05 &   6.33 & 0 & +133 & -80.9 & 155 & noBH-18\\
   100 &   300 &      0.638 &  2.95  & 18.62 &      4219 &    4.05 &   2.16 &   6.65 & 0 & +172 & -78.3 & 189 & noBH-19\\
   200 &   9.5 &       1.28 &  11.7  & 10.79 &      1062 &    8.76 &  0.501 &   11.7 & 0 & -31.5 & -53.2 & 61.8 & noBH-10\\
   200 &    10 &       1.28 &  11.5  & 11.65 &      1675 &    9.03 &  0.812 &     12 & 0 & +8.11 & -51.7 & 52.4 & noBH-6\\
   200 &    30 &       1.28 &  7.98  &  2.35 &     10320 &    9.52 &   6.49 &   12.7 & 0 & +566 & -49.1 & 568 & noBH-20\\
   300 &   9.5 &       1.91 &  13.4  & 11.11 &      1065 &    13.5 &  0.508 &   17.5 & 0.321 & -31.3 & -27.7 & 41.8 & noBH-5\\
   300 & 9.625 &       1.91 &  13.3  & 11.38 &      1349 &    12.2 &   0.67 &     16 & 0.244 & -13 & -34.7 & 37.1 & noBH-4\\
   300 &  9.75 &       1.91 &  13.3  & 11.45 &      1500 &    12.2 &  0.736 &   16.1 & 0.222 & -3.23 & -35 & 35.1 & noBH-3\\
   300 &    10 &       1.91 &  13.2  & 12.07 &      2316 &    12.1 &    1.2 &   16.1 & 0.136 & +49.5 & -35.5 & 60.9 & noBH-9\\
   350 & 9.625 &       2.23 &    14  & 11.37 &      1136 &    19.6 &  0.583 &   25.3 & 0 & -26.7 & +4.84 & 27.1 & noBH-1\\
   350 &  9.75 &       2.23 &    14  & 11.63 &      1864 &    13.7 &  0.966 &   17.9 & 0 & +20.3 & -26.7 & 33.6 & noBH-2\\
   400 &   9.5 &       2.55 &  14.7  & 11.14 &        81 &    22.5 & 0.0436 &   33.5 & 0 & -94.8 & +20.2 & 96.9 & noBH-12\\
   400 & 9.625 &       2.55 &  14.7  & 11.40 &      1105 &    27.1 &  0.539 &   34.7 & 0 & -28.7 & +45.1 & 53.4 & noBH-7\\
   400 &  9.75 &       2.55 &  14.6  & 11.79 &      2516 &    14.3 &   1.28 &   19.1 & 0 & +62.3 & -23.7 & 66.7 & noBH-11\\
   400 &    10 &       2.55 &  14.5  & 12.16 &      3237 &    14.1 &    1.7 &   19.1 & 0 & +109 & -24.4 & 112 & noBH-16\\
   500 &   9.5 &       3.19 &  15.9  & 11.14 &       112 &      24 & 0.0573 &   30.8 & 0 & -92.8 & +28.2 & 97 & noBH-13\\
   500 &  9.75 &       3.19 &  15.7  & 11.66 &      2429 &    18.7 &   1.24 &   24.6 & 0 & +56.7 & +0.0658 & 56.7 & noBH-8\\
   500 &    10 &       3.19 &  15.6  & 11.92 &      3650 &    15.8 &   1.86 &   21.1 & 0 & +136 & -15.3 & 136 & noBH-17\\
 \hline
 \hline
 \caption{As in Table 1 but now for the 20 noBH models.}
\end{longtable}
\end{center}
\end{spacing}
\end{landscape}

\clearpage\newpage
%STARTmethods
\section*{Methods}
\noindent{\bf Milky Way model and orbit of Pal 5.}
We adopt a time-independent,  axisymmetric  Milky Way, consisting of a dark matter (DM) halo, a disc and a bulge. We derive the Milky Way parameters and the orbit of Pal 5 from a fit to the Pal 5 stream. This stream fit is nearly identical to that in Erkal et al.\cite{2017MNRAS.470...60E} which used the \texttt{MWPotential2014} potential of Bovy\cite{2015ApJS..216...29B}, but we use updated priors based on more recent measurements of distance\cite{2019AJ....158..223P} and proper motion\cite{2019MNRAS.484.2832V} of Pal 5 and the  distance to the Galactic center\cite{2019A&A...625L..10G}. The fit treats Pal 5's present-day distance, radial velocity, proper motions, the distance to the Galactic center, and the DM-halo mass and scale radius as free parameters. 

The DM halo is described by a spherical Navarro-Frenk-White (NFW) profile\cite{1996ApJ...462..563N}, with potential 
\begin{equation}
\Phi_{\rm halo}(R_{\rm G}) = -\frac{GM_{\rm NFW}}{R_{\rm G}}\ln\left(1+\frac{R_{\rm G}}{a_{\rm NFW}}\right),
\end{equation}
where $R_{\rm G}$ is the Galactocentric radius, $M_{\rm NFW}=4.38713\times10^{11}~M_\odot$  is a mass scale and $a_{\rm NFW}=16.043~$kpc is the scale radius. For these parameters the virial mass and concentration are $M_{\rm vir}=8.127\times10^{11}~M_\odot$ and $c=15.3$, respectively. We note that these parameters are close to those of Bovy\cite{2015ApJS..216...29B}. The disc potential is that of an axisymmetric Miyamoto-Nagai disc\cite{1975PASJ...27..533M}
\begin{equation}
\Phi_{\rm disc}(R_{\rm G}) = - \frac{GM_{\rm MN}}{\left[X^2 + Y^2 + \left(a_{\rm MN} + \sqrt{Z^2 + b^2}\right)^2\right]^{1/2}},
\end{equation}
where $M_{\rm MN} = 6.8\times10^{10}~M_\odot$ is the mass of the disc,  $a_{\rm MN}=3~$kpc is the disc scale length and $b=0.28~$kpc is the scale height and $X,Y,Z$ are  Cartesian Galactocentric coordinates  (that is $R_{\rm G}^2 = X^2 + Y^2 + Z^2$). The bulge is described by a Hernquist potential\cite{1990ApJ...356..359H}
\begin{equation}
\Phi_{\rm bulge}(R_{\rm G})  = - \frac{GM_{\rm H}}{R_{\rm G} + a_{\rm H}}, 
\end{equation}
with $M_{\rm H} = 5\times10^{9}~M_\odot$ the bulge mass and $a_{\rm H}=0.5~$kpc its length scale. The bulge model is slightly different from what was used by Erkal et al.\cite{2017MNRAS.470...60E}, but within the orbit of Pal 5 the total enclosed mass of both bulge models is the same.

Cartesian Galactocentric phase-space coordinates of Pal 5 in the potential described above were given by the fit: $[5.733, 0.2069,14.34]~$kpc and $[-41.33, -111.8, -16.85]~$km/s. The Sun is found to be at  $[-8.182, 0, 0]~$kpc, with a  velocity of  $[11.1, 245.7, 7.3]~$km/s, which puts Pal 5 at a distance from the Sun of 19.98 kpc. To find the initial position and velocity of Pal 5's orbit, we flipped the sign of the velocity vector and integrated the orbit backward in time. We adopt an age of 11.5 Gyr\cite{2002AAS...201.0711M,2011ApJ...738...74D,2021AJ....161...12X} and fix this for all models. The typical uncertainty on the age determinations is $\sim1$ Gyr and varying the age would somewhat increase the uncertainty on our initial parameters, but not the parameters of our best model.  The initial Galactocentric position and velocity  of Pal 5 are $[1.212, 10.08,  2.744]$~kpc and $[ 43.16, -162.6, 147.0]$~km/s, respectively.

%%%%
\noindent{\bf Cluster initial conditions and $N$-body code.}
The initial positions and velocities of the stars were sampled from an isotropic Plummer model\cite{1911MNRAS..71..460P}, truncated at 20 scale radii. Initial stellar masses were sampled from a Kroupa IMF\cite{2001MNRAS.322..231K} in the range $0.1-100~M_\odot$ and a metallicity of $Z=6\times10^{-4}$ was adopted, that is ${\rm [Fe/H]}\simeq-1.4$\cite{2002AJ....123.1502S}. All simulations were run with the direct (that is, no softening) $N$-body code {\sc nbody6++gpu}\cite{1999PASP..111.1333A, 2003gnbs.book.....A,2015MNRAS.450.4070W}, which deploys a 4$^{\rm th}$-order Hermite integrator with an Ahmad-Cohen neighbour scheme\cite{1973JCoPh..12..389A, 1992PASJ...44..141M}. It has recipes for stellar and binary evolution\cite{2000MNRAS.315..543H, 2002MNRAS.329..897H}, with recent updates for BH masses and kicks\cite{2020A&A...639A..41B} and it deals with close encounters with Kustaanheimo-Stiefel (KS) regularisation. We  use the Graphics Processing Unit (GPU)-enabled\cite{2012MNRAS.424..545N} version and the simulations were run on a server with GeForce RTX 2080 Ti GPUs at ICCUB.  A few modifications to the code were made for this project. The singular  isothermal halo was replaced by the NFW halo, with the force and force derivatives derived from equation~(1). Stars were stripped from the simulation when they reached a distance of 40 times the instantaneous half-mass radius of the bound particles and for each escaper the time, position and velocity in the Galactic frame was stored. The escapers were then integrated as tracer particles in the Galactic potential until 11.5 Gyr with a separate integrator to construct the tidal tails. We did not include the contribution of the cluster to the equation of motion of the tail stars. Near the end of the simulation the fractional contribution of the cluster to the total force is approximately $2\times10^{-5}$ for stars that just left the cluster, and smaller for stars at larger distances,  justifying this assumption. All models were evolved until complete dissolution, and for the models that dissolved after 11.5 Gyr,  a snapshot was saved at exactly 11.5 Gyr (that is, when the cluster is at the position of Pal 5 today).

\noindent{\bf Black hole recipes.} 
The fast evolution codes {\sc sse}\cite{2000MNRAS.315..543H} for stars and {\sc bse}\cite{2002MNRAS.329..897H} for binaries in {\sc nbody6++gpu} were recently  modified\cite{2020A&A...639A..41B} with updated prescriptions for wind mass loss, compact object (that is neutron stars and BHs) formation and supernova kicks\cite{2012ApJ...749...91F}.  
We adopt here the rapid supernova mechanism in which the explosion is assumed to occur within the first 250ms after bounce\cite{2012ApJ...749...91F}, which corresponds to \texttt{nsflag=3} in {\sc sse/bse}.
 The BH natal kick velocities are drawn from a Maxwellian with dispersion $\sigma=265$~km/s\cite{2005MNRAS.360..974H}, and in the wBH models they are subsequently lowered by the amount of fallback such that momentum is conserved (that is, \texttt{kmech = 1}).
As a result,  63\%(73\%) of the number(mass) of BHs does not receive a kick for the IMF and metallicity we used. The BHs that receive a kick form from stars with zero age mean sequence  (ZAMS) masses in the range $23-34~M_\odot$, resulting in BH masses in the range $6-26~M_\odot$, and
because of the low escape velocities of our model clusters (10-20~km/s) they are almost all lost. As a result, the lowest mass BH in the cluster just after supernovae and kicks has a mass of $13~M_\odot$ and the average BH mass is $21~M_\odot$. 

In natal kick models that consider the effect  of fallback of mass on the BH\cite{2008ApJS..174..223B}, the exact fraction of BHs that do not receive a kick  depends on the  prescription for compact-object formation. For our IMF and metallicity we find  for
 the {\sc StarTrack}, rapid, and delayed explosion  mechanisms described in section 4 of Fryer et al.\cite{2012ApJ...749...91F}, that about $83\%$, $73\%$ and $56\%$  of the total BH mass is in BHs that form without a natal kick, respectively.
 Our result of $73\%$ is therefore an intermediate value. A lower(higher) initial $f_{\rm BH}$  requires a lower(higher) initial cluster density to end up with the same cluster properties at the present day.

\noindent{\bf Pal 5 parameters.} To find the number of observed stars ($N_{\rm cluster}$) and half-light radius ($R_{\rm eff}$) of Pal 5, we fit King models\cite{1966AJ.....71...64K} to the density profile of figure 10 of  Ibata et al.\cite{2013MNRAS.428.3648I} using  {\sc limepy}\cite{2015MNRAS.454..576G} and the  Markov Chain Monte Carlo (MCMC) code {\sc emcee}\cite{2013PASP..125..306F}. We fit for the dimensionless central potential: $W_0=4.6\pm0.3$, the number of observed stars: $N_{\rm cluster}=1,550\pm40$, the half-mass radius: $r_{\rm h} = 4.3\pm0.1~$arcmin and the background:  $0.07\pm0.01$~arcmin$^{-2}$.
For these parameters $R_{\rm eff} = 3.21\pm0.06$ arcmin, which for the adopted distance to Pal 5 of 19.98 kpc corresponds to $R_{\rm eff} = 18.7\pm0.4$ pc.

\noindent{\bf Data analysis.} For each simulation, snapshots were saved  approximately every 50 Myr. For each snapshot, we  find the stars and remnants that are energetically bound to the cluster, those that  have a specific energy $E_i= 0.5v_i^2 +\phi_i<0$, where $v_i$ is the velocity in the cluster's center of mass frame and $\phi_i$ is the potential due to the mass of  the other bound stars, which we determine iteratively. From the  bound particles we determine the total  cluster mass $M$, the mass of the BH population $M_{\rm BH}$, the half-mass radius $r_{\rm h}$, $N_{\rm cluster}$ and $R_{\rm eff}$.  

To compare the $N$-body models to observations, we use isochrones to convert masses of stars in different evolutionary phases to magnitudes.  We use Sloan Digital Sky Survey (SDSS) $g$-band magnitudes from MESA Isochrones and Stellar Tracks (MIST) isochrones\cite{2016ApJ...823..102C, 2016ApJS..222....8D} for 11.5 Gyr, $Z=6\times10^{-4}$  (${\rm [Fe/H]}=-1.4$), ${\rm [}\alpha/{\rm Fe]}=0$ and rotational velocities of 0.4 times critical. 
For the observations shown in Figure 1, Canada-France-Hawaii Telescope (CFHT) photometry was used\cite{2016ApJ...819....1I} which is a slightly different photometric system than SDSS. A color transformation is provided by Ibata et al.\cite{2016ApJ...819....1I}. For the magnitude limits we adopt here, the colors of Pal 5 stars vary between $0.3\lesssim(g-r)_{\rm CHFT}\lesssim0.5$ in the relevant magnitude range, resulting in corrections in the $g$-band between $0.07-0.11$. We therefore adopt $g_{\rm SDSS} = g_{\rm CFHT} + 0.1$ to convert between the two systems. For the  observed number density profile, a magnitude range of $19<g_{\rm CFHT}<23$ was used by Ibata et al.\cite{2017ApJ...842..120I}. We apply the corresponding  magnitude range $19.1 < g_{\rm SDSS} < 23.1$ and combined with the adopted distance (DM=16.5 mag) this implies a mass range of main sequence stars of $0.625- 0.815~M_\odot$, which is what we use to select stars in the $N$-body model to construct the surface density profile. 
For the stream we selected stars with $20.3<g_{\rm CFHT}<23.5~$mag (that is $20.4<g_{\rm SDSS}<23.6$) as in Ibata et al.\cite{2016ApJ...819....1I}. This magnitude cut  implies mass limits that depend on the distance  of stars in the tidal tails.

\noindent{\bf Finding the best model.} We varied the initial number of stars $N_0$ and the initial half-mass density $\rho_{\rm h0}$ and try to reproduce $N_{\rm cluster}$ and $R_{\rm eff}$.  For the wBH models, we first ran a coarse grid of models with $N_0= [50k, 100k, 200k]$ and $\rho_{\rm h0} = [30, 100, 300, 10^3]~M_\odot/{\rm pc}^3$. We ran each combination of $N_0$ and $\rho_{\rm h0}$, apart from $N_0=200k$ and $\rho_{\rm h0}=10^3~M_\odot/{\rm pc}^3$. 
For each model we computed the fractional differences between the $N$-body results and the observations for $N_{\rm cluster}$ and $R_{\rm eff}$ at 11.5 Gyr: $\delta_N$ and $\delta_R$, respectively, and then define the best model as the one for which $\delta = \sqrt{\delta^2_N + \delta_R^2}$ is lowest. For each model we also find the time to reach $N_{\rm cluster}=1,550$ (that is, $T_{\rm 1,550}$). We introduce $T_{1,550}$ to establish a goodness of fit  for models that dissolve before 11.5 Gyr. Model IDs are increasing with the value of $\delta$ and for clusters that dissolved the ID is in order of increasing difference between $T_{1,550}$  and 11.5 Gyr. 
The model with the smallest $\delta$ in the coarse grid has  $N_0 = 200k, \rho_{\rm h0} = 100~M_\odot/{\rm pc}^3$ and reproduces the two observables within 10\%. From the  coarse grid we estimate that a slightly larger $N_0$ and lower $\rho_{\rm h0}$ would reduce the difference. 
We then ran a finer grid with 6 more models with $N_0$ in the range $210k-225k$ and $\rho_{\rm h0}$ in the range $70-80~M_\odot/{\rm pc}^3$. The model with the smallest $\delta$, wBH-1, has $N_0 = 210k$ and $\rho_{\rm h0} = 80~M_\odot/{\rm pc}^3$. It reproduces both observables within 3\% and it has a present-day mass of $9.5\times10^3~M_\odot$, $r_{\rm h}=18.8~$pc and $f_{\rm BH} = 0.22$. For the noBH models, we first ran five models with $N_0 = [100k, 200k, 300k, 400k, 500k]$ and $\rho_{\rm h0} = 10~M_\odot/{\rm pc}^3$, two models with $N_0 = [100k, 200k]$ and $\rho_{\rm h0} = 30~M_\odot/{\rm pc}^3$ and two models with $N_0 = 100k$ and $\rho_{\rm h0} = [100, 300]~M_\odot/{\rm pc}^3$. Models with $\rho_{\rm h0} = 10~M_\odot/{\rm pc}^3$ gave results similar to the observations. 
We then ran an additional 11 models with densities $\rho_{\rm h0} \lesssim 10~M_\odot/{\rm pc}^3$ and $200k\le N_0\le 500k$ and found that the model with the lowest $\delta$ has $N_0=350k$ and $\rho_{\rm h0} = 9.625~M_\odot/{\rm pc}^3$.
 All model results are summarised in Tables 1 and 2.

\noindent{\bf Rate of growth of the tidal tails and  their visibility.}
From the escaping stars we find that half of the observable stars in the tails of wBH-1 were ejected in the final 3 Gyr.  
{To estimate the number of MW field stars in Figure 4 that share the same locus in 
color-magnitude as Pal 5, we use the CFHT data and color-magnitude selection 
criteria from Ibata et al.\cite{2016ApJ...819....1I}. Additionally, we select only stars more than 0.4 deg 
away from the best-fit stream track from Erkal et al.\cite{2017MNRAS.470...60E} and adopt their magnitude limits of $20.0<g_{\rm CFHT}<23.5~$mag. This sample is dominated 
by MW field stars and has an average density of 0.142 stars/arcmin$^2$. 
Note that Pal 5 stream runs roughly at a constant $b$ in the region explored 
in Figure~5, thus variations in the MW field density with position are negligible.}

\noindent{\bf Predictions for observations.} To estimate whether the BH population can be detected from the kinematics, we look at the velocity dispersion of the wBH-1 model. Within  $R_{\rm eff}$ ($\sim3.2$ arcmin), there are 40 giant stars, with a line-of-sight velocity dispersion of $0.69\pm0.09~$km/s. Because our wBH-1 model has a steeper mass function than Pal 5, the mass of wBH-1 is likely too high. We therefore scale our model results to a conservative mass of $5\times10^3~M_\odot$\cite{2017ApJ...842..120I} (excluding BHs) and  $r_{\rm h}=25~$pc. This reduces the predicted dispersion of the giants to $0.58\pm0.06~$km/s. In the noBH-1 model we find 51 giants within $R_{\rm eff}$, which have a (scaled) line-of-sight dispersion of $0.42\pm0.04~$km/s.
We then fit King models\cite{1966AJ.....71...64K} and a constant background using {\sc limepy}\cite{2015MNRAS.454..576G} to the number density profile of bright main sequence stars shown in Figure 1(a). In Figure 6(a,b) we show the results. For wBH-1 we find a dimensionless central concentration of $W_0 = 5.9\pm0.2$, $r_{\rm h} = 27.7\pm{0.8}$~pc, which results in $R_{\rm eff} = 20.8\pm0.6$~pc. For noBH-1 we find $W_0 = 5.2\pm0.3$, $r_{\rm h} = 31.5\pm{1.1}$~pc, which results in $R_{\rm eff} = 23.9\pm0.4$~pc. 
We then derive the line-of-sight velocity dispersion of the King models by adopting a total mass (excluding the BHs) of: $5\times10^3~M_\odot$ and  $r_{\rm h}=25~$pc. The resulting velocity dispersion profiles are shown in Figure 6(c,d). The central dispersion of the King models is $\sim0.4$~km/s. The dispersion of the giants in the noBH-1 model agrees with this, which means that the derived dispersion provides an accurate measure of the total mass when assuming that the surface density traces the total mass.
For the wBH-1 model, however, the dispersion of the giants is about 50\% (or 200~m/s) higher. 
A small part of this difference can be explained by the fact that wBH-1 is a factor of $1/(1-0.22)\simeq1.3$ more massive, because of the BHs. This higher mass increases the dispersion by a factor of  $\sqrt{1.3}\simeq 1.14$. The additional factor of 1.3 needed to explain the dispersion of the giants is because the BHs are centrally concentrated, inflating the central dispersion more than in the mass follows light assumption.  Although it is challenging to find a 200~m/s velocity difference, it is feasible with existing high-resolution spectographs ($R\sim20\,000$) on $8$m-class telescopes with a multi-epoch observing strategy.  
Baumgardt \& Hilker\cite{2018MNRAS.478.1520B} present a compilation of line-of-sight velocities of Milky Way GCs. There are 32 stars in their Pal 5 sample, and the dispersion of the inner 15 stars is $0.55^{+0.15}_{-0.11}~$km/s. From a comparison of $N$-body models to the kinematics and surface brightness profiles the authors derive a central dispersion of 0.6~km/s. For such a low dispersion the orbital motions of binary stars are important. The authors have repeat observations for about half of their stars, and they reject stars with large velocity variations. Their result supports the BH hypothesis, but a more thorough analysis of the binary content is desirable.  
Pal~5 has approximately $50(100)$ stars brighter than $g<19(20)~$mag within 3~arcmin from the center for which (additional) line-of-sight velocities can be obtained. For individual measurement errors of 300 m/s and a velocity dispersion of 400~m/s, the estimated uncertainty in the velocity dispersion for 50(100) stars is 63(43) m/s, that is smaller than the difference between wBH-1 and noBH-1. Similar uncertainties can be obtained even when simultaneously fitting on the velocity dispersion and the properties of binary stars\cite{2012A&A...547A..35C}. 

The BH scenario also make a critical prediction for the properties of the binary stars.
Soft binaries are ionised when they interact with stars\cite{1975MNRAS.173..729H} and the  binaries with the lowest binding energy are therefore  indicators of the most energetic cluster members, including invisible remnants. From wBH-1 we find that the average kinetic energy of the BHs is $\langle K_{\rm BH}\rangle = \langle 0.5mv^2\rangle \simeq 10~M_\odot({\rm km/s})^2$, while for all the other stars, white dwarfs and neutron stars we find $\langle K_*\rangle \simeq 0.3~M_\odot({\rm km/s})^2$. Adopting circular orbits and equal-mass binary components of $0.8~M_\odot$ (that is, the mass of giants for which we can obtain velocities) we find that the orbital period is capped at $P_{\rm max}\simeq2.3\times10^4(120)~$yr because of interactions with stars(BHs).  However, finding no binaries with $P>120~$yr does not confirm the presence of BHs, because the cluster was initially more massive and  compact, such that soft binaries had shorter periods. From the initial mass and half-mass radius of wBH-1(noBH-1) we find that the initial $\langle K\rangle$ was a factor of $45(30)$ higher, implying that in the BH case, the maximum binary period is $P_{\rm max} \simeq0.5~$year, while in the noBH hypothesis the maximum period is $P_{\rm max} \simeq 140~$yr. This suggests that the presence of binaries with $0.5~\lesssim P/{\rm yr}\lesssim 140~$ would be an argument against the presence of a BH population, while the absence of such binaries combined with more detailed predictions from $N$-body simulations with  primordial binaries could be  used as support  for the BH case. For $P=[1,10,100]$~yr, the orbital velocities are $[12,5.8,2.7]~$km/s, hence these binaries would be easily detectable with a baseline of weeks/months and moderate spectral resolution.
In addition, multi-epoch kinematics  serves to look for stars with a (detached) BH binary companion such as found in NGC3201\cite{2018MNRAS.475L..15G, 2019A&A...632A...3G}. Although dynamical formation of BH binaries is more common, a BH with stellar companion can form in an exchange interaction\cite{2018ApJ...855L..15K}. 

We note that several  studies have put forward observational signatures of BH populations in clusters by using dynamical models. These studies used either the degree of mass segregation\cite{2016MNRAS.462.2333P, 2016ApJ...833..252A,2018ApJ...864...13W,2020ApJ...898..162W} or other observational properties of GCs\cite{2018MNRAS.478.1844A} to make predictions for the size of the BH populations in a subset of  Milky Way GCs. None of these studies included Pal 5, so we can not make a comparison. We can check whether these methods would be able to infer the correct BH population given our model properties. From the scaling between $f_{\rm BH}$ and the ratio of core over half-light radius ($R_{\rm core}/R_{\rm eff}$) shown in figure 7 of Weatherford et al.\cite{2020ApJ...898..162W} 
and the ratio $R_{\rm core}/R_{\rm eff} \simeq 0.5$ in wBH-1, we would infer that Pal 5 has a BH population of (only) $f_{\rm BH} \simeq 0.3-1\%$. The reason the inferred $f_{\rm BH}$ from their relation is so different could be because $R_{\rm core}/R_{\rm eff}$ saturates or goes down again for vary large $f_{\rm BH}$ where they have no models, or because of a systematic difference in $R_{\rm core}$ between the Monte Carlo and $N$-body methods\cite{2016MNRAS.463.2109R,2018ComAC...5....5R}. Using the fitted relations provided by Askar et al.\cite{2018MNRAS.478.1844A} (see their Table A1) between the global properties of stellar-mass BHs and the observational properties of their host globular cluster, they find that about 50\% of the mass in Pal 5 could be in stellar-mass BHs. 

\noindent{\bf Microlensing event rate.} For a BH mass of $17~M_\odot$ at a distance of 20 kpc, the Einstein angle is $\theta_{\rm E} \simeq2.6~$mas. In wBH-1 there are 124 BHs, mostly within $R_{\rm eff}\simeq3~$arcmin, resulting in a surface density of $\Sigma_{\rm BH}\simeq1.2\times10^{-9}~$mas$^{-2}$. The proper motion of Pal~5 is  $\mu\simeq3.2~$mas/yr, such that the event rate for a single background source (that is, an active galactic nucleus) is $\theta_{\rm E}\Sigma_{\rm BH}\mu\simeq10^{-8}$/yr. In the recent active galactic nuclei catalogue from the Gaia and WISE surveys\cite{2019MNRAS.489.4741S} we find 2 active galactic nuclei within $R_{\rm eff}$ of Pal 5, too low to get to a reasonable event rate. The event rate would  be larger if we also consider background halo stars, but even with the estimated density of background galaxies in LSST  ($\sim50/{\rm arcmin}^2$), the lensing event rate would be too low ($\sim10^{-5}/$yr).
%ENDmethods

%%%%%%%%%%%%%%%%%%%%%%%%%%%%%%%%%%%%%%%%%%%%%%%%%%%%%%%%%%%%
%%%%%%%%%%%%%%%%%%%%%%% SUPPLEMETARY INFORMATION%%%%%%%%%%%%%%%%%%%%
%%%%%%%%%%%%%%%%%%%%%%%%%%%%%%%%%%%%%%%%%%%%%%%%%%%%%%%%%%%%
\clearpage\newpage
\noindent{\Large Supplementary Information}

\section*{Sensitivity to initial conditions}
We determine how sensitive the final parameters are to variations in the initial parameters. For a quantity $A$, we express the difference between its value for model $i$ ($A^i$) and the best model ($A^1$) as $\Delta \log A = \log(A^i/A^1)$, where $A$ can be an initial property ($N_0$ or $\rho_{\rm h0}$) or an observable property. For the latter we use $N_{\rm cluster}$ and the number density within the half-light radius: $\rho_{\rm eff} = 3N_{\rm cluster}/(8\pi R_{\rm eff}^3)$. We can write the variation in the final properties in terms of the initial properties as
\begin{equation}
\begin{pmatrix}
\Delta \log N_{\rm cluster} \\
\Delta \log \rho_{\rm eff}
\end{pmatrix}
=
\Sigma
\begin{pmatrix}
\Delta \log N_0 \\
\Delta \log \rho_{\rm h0}
\end{pmatrix}.
\end{equation}
Here $\Sigma$ is a matrix that contains the constants that relate variations in initial parameter to variations in the final parameters.
We find the four elements of $\Sigma$ from the two models that are nearest to the observations  (that is, wBH-2, wBH-3  and noBH-2, noBH-3)
\begin{equation}
\Sigma_{\rm wBH} = 
\begin{pmatrix}
-5.74 & -1.56\\
-7.19 & -1.57
\end{pmatrix}, 
\Sigma_{\rm noBH} = 
\begin{pmatrix}
1.47 & 38.4\\
-0.71 & 122
\end{pmatrix}.
\end{equation}
Absolute values of 1 mean that a fractional change in an initial parameter leads to the same fractional change in the final parameter. The absolute values $>1$ in the left column of $\Sigma_{\rm wBH}$ mean that the final properties are most sensitive to $N_0$, which is because of the collisional nature of the wBH models. The $>1$ value in the right column of $\Sigma_{\rm noBH}$ show that the final parameters are most sensitive to the initial density, which is because of the collisionless nature of these models. Taking the inverse of the $\Sigma$ matrices, and assuming small variations, that is, $\Delta \log A \simeq\log (1+\epsilon_A)\propto\epsilon_A$, with $\epsilon_A << 1$,   we can write
\begin{equation}
\begin{pmatrix}
\epsilon_{N_0} \\
\epsilon_{\rho_{\rm h0}}
\end{pmatrix}_{\rm wBH}
\simeq
\begin{pmatrix}
0.711 & -0.707\\
-3.26 & 2.60
\end{pmatrix}
\begin{pmatrix}
\epsilon_{N_{\rm cluster}} \\
\epsilon_{\rho_{\rm eff}}
\end{pmatrix}
\end{equation}
and
\begin{equation}
\begin{pmatrix}
\epsilon_{N_0} \\
\epsilon_{\rho_{\rm h0}}
\end{pmatrix}_{\rm noBH}
\simeq
\begin{pmatrix}
0.590 & -0.186\\
0.00345 & 0.00713
\end{pmatrix} 
\begin{pmatrix}
\epsilon_{N_{\rm cluster}} \\
\epsilon_{\rho_{\rm eff}}
\end{pmatrix}\ .
\label{eq:eps0}
\end{equation}
This shows that the level of fine-tuning to obtain the correct $N_0$ is similar in both models, albeit more sensitive to variations in $\rho_{\rm eff}$ for the noBH models. However, we find different behaviour for  the initial density, $\rho_{\rm h0}$: for the wBH models, variations in $N_{\rm cluster}$ and $\rho_{\rm eff}$ allow for larger variations in $\rho_{\rm h0}$, meaning that a relatively large range of initial densities can contribute to the error bars on the present-day properties. The results of the noBH models are extremely sensitive to the initial density, because variations in the observed properties correspond to variations of less than a  per cent in the initial density.
We can also estimate what fraction of the parameter space of the initial conditions is covered by the uncertainties in $N_0$ and $\rho_{\rm h0}$. We assume that the initial cluster properties are sampled from power-law distributions with indices $-2$ for $N_0$\cite{2019ARA&A..57..227K_} and $-1$ for $\rho_{\rm h0}$\cite{2005ApJ...623..650K_} in the ranges $10^5\le N_0\le 5\times10^6$ and $1\le \rho_{\rm h0}/(M_\odot\,{\rm pc}^{-3})\le 10^4$. Then we find that the initial conditions of wBH-1(noBH-1)  that contribute to the error circle cover a fraction 1/200(1/$3.6\times10^5$) of the initial conditions. Given that the Milky Way has $\sim150$ GCs, this exercise shows that finding Pal 5 is probable in the wBH scenario, while the probability in the noBH scenario is $10^{-3}$. 
\begin{figure}[!h]
\includegraphics[width=14cm]{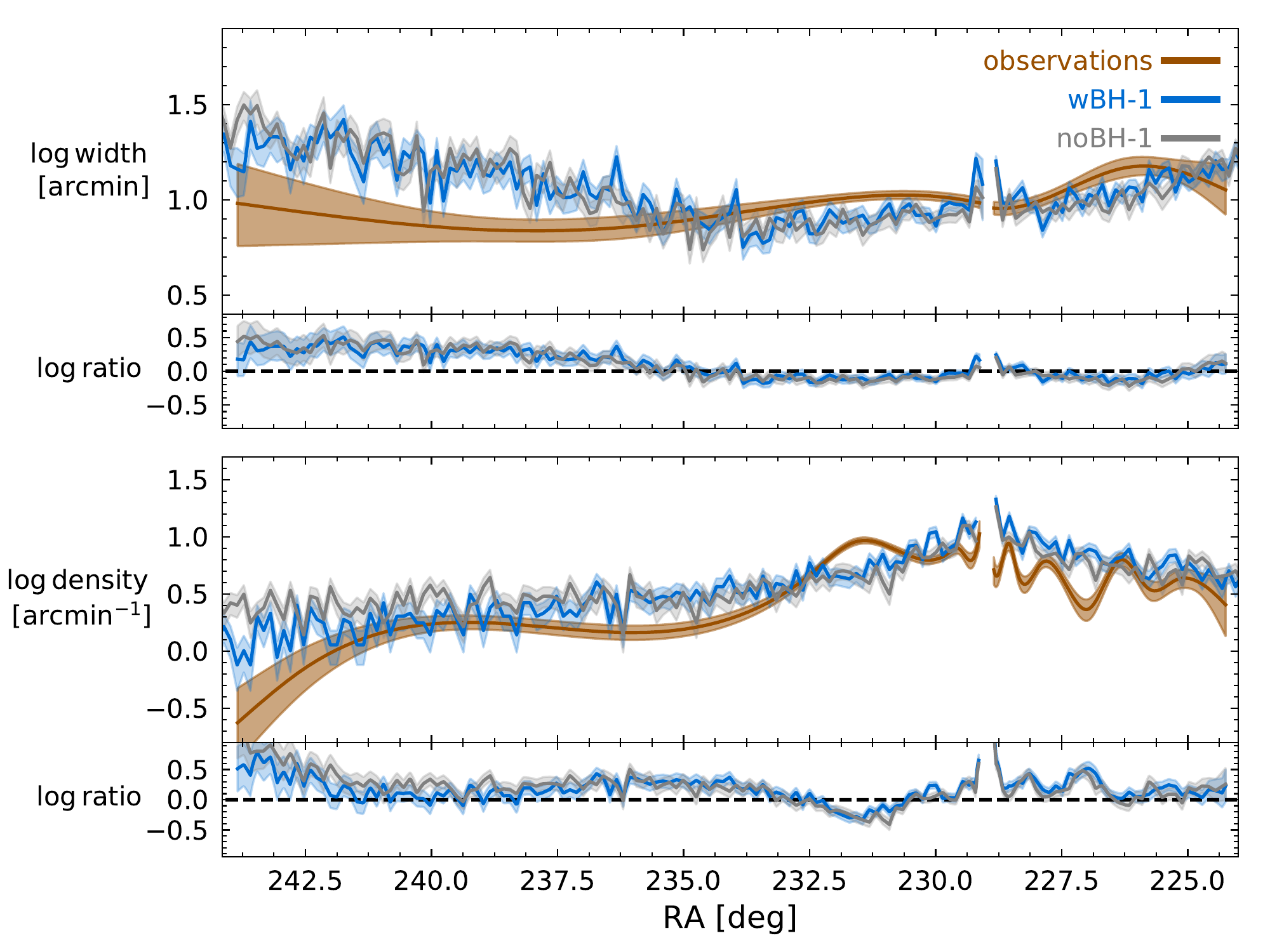}\\
{Supplementary Figure 1: {\bf Comparison of stream properties.} Comparison between stream properties of wBH-1 and noBH-1 and the observed stream from Erkal et al.\cite{2017MNRAS.470...60E_}. The stream width (top) of both $N$-body models is similar over the range included in the observations and shows some systematic deviations from the observed width. The density profile (bottom) of the $N$-body models is smoother than the observed profile, which shows signatures of over/under-densities. The decline in the density of the trailing arm (large RA) is faster in wBH-1 than in noBH-1, which agrees more with the observed decline.  Shaded regions indicate the 67\% confidence intervals.}
\end{figure}

\clearpage\newpage
\begin{spacing}{1.0}
\begin{center}
\small
\begin{longtable}{ l c c | l c c } 
 \hline\hline
  \multicolumn{3}{c}{compact GCs} &  \multicolumn{3}{c}{`fluffy' GCs}\\
  name & distance [kpc] & tail data & name & distance [kpc] & tail data\\\hline
NGC~3201 & 4.9 & long tidal tails\cite{2020arXiv201205245I_,2020arXiv201014381P_}  & NGC~288 &8.9 & long tidal tails\cite{2019MNRAS.484L.114K_,2018ApJ...862..114S_,2020arXiv201205245I_} \\
NGC~6205 & 7.1 &  tidal tails\cite{2000A&A...359..907L_} & Pal~1 &11.1 & tidal tails\cite{2010MNRAS.408L..66N_} \\
NGC~6341  &8.3 &  long tidal tails\cite{2020arXiv201205245I_} & NGC~6101 & 15.4 &  - \\
NGC~362   & 8.6 & tidal feature\cite{2019MNRAS.486.1667C_}& NGC~5466 &16.0 &  tidal tails\cite{2006ApJ...637L..29B_,2016MNRAS.463.1759B_} \\
NGC~6779 & 9.4 & - & NGC~5053 & 17.4 & tidal feature\cite{2010A&A...522A..71J_} \\
NGC~2808 & 9.6 & - & IC~4499 & 18.8 & - \\
NGC~5272 & 10.2 & -    & BH~176 & 18.9 & - \\
NGC~4590 & 10.3 & long tidal tails\cite{2020arXiv201205245I_}& Pal~12 & 19.0 & tidal tails\cite{2000A&A...359..907L_} \\
NGC~7078 & 10.4 & -   & NGC~6426 &20.6 & - \\
NGC~2298 & 10.8 & tidal features\cite{2011MNRAS.416..393B_}& Rup~106 &21.2 &  -\\
NGC~7089 & 11.5 &-  & ESO~280 &21.4 &  - \\
NGC~5286 & 11.7 & -& Ter~7 & 22.8 & - \\
NGC~1851 &12.1 & long tidal tails\cite{2018ApJ...862..114S_, 2020arXiv201205245I_}& Pal~5 & 23.2 & long tidal tails\cite{2001ApJ...548L.165O_, 2020ApJ...889...70B_,2020arXiv201205245I_} \\
NGC~1904 & 12.9 & tidal tails\cite{2018ApJ...862..114S_}& IC~1257 & 25.0 & - \\
NGC~6934 &15.6 &  -& Pal~13 & 26.0 & long tidal tails\cite{2020AJ....160..244S_} \\
NGC~1261 &16.3 &  long tidal tails\cite{2018ApJ...862..114S_,2020arXiv201205245I_}& Ter~8 & 26.3 & - \\
NGC~5024 & 17.9 & -     & NGC~7492 & 26.3 & tidal tails\cite{2017ApJ...841L..23N_} \\
NGC~6981 &17.0 &   - & Arp~2 &28.6 &  - \\
NGC~4147 & 19.4 & tidal feature\cite{2010A&A...522A..71J_} & AM~4 & 32.2 & - \\
NGC~6864 & 20.9 & -    & Pyxis &39.4 &  - \\
NGC~5634 & 25.2 &  -   & Pal~15 & 45.1 & tidal tails\cite{2017ApJ...840L..25M_} \\
Pal~2          &27.2 & -    & Pal~14 & 76.5 &  tidal tails\cite{2011ApJ...726...47S_} \\
NGC~6229 & 30.5 & -  & Eridanus & 90.1 & tidal tails\cite{2017ApJ...840L..25M_} \\
NGC~5824 & 32.1 & -  & Pal~3 & 92.5 & tidal feature\cite{2003AJ....126..803S_} \\
NGC~5694 & 35.0 & -  & Pal~4 & 108.7& tidal feature\cite{2003AJ....126..803S_} \\
NGC~7006 & 41.2 & -     & AM~1 & 123.3 & - \\   
  \hline\hline
\caption*{Supplementary Table 1: Summary of results of stream searches for GCs at distances $>8$ kpc from the Galactic center. The classification `compact' (left) and `fluffy' (right) is from Baumgardt et al.\cite{2010MNRAS.401.1832B_}. All clusters are sorted in distance from the Sun. Among the compact clusters, no tidal tails nor tidal features were found for clusters that are more than 20 kpc away, while they were found for about half of the fluffy clusters beyond 20 kpc. }
\end{longtable}
\end{center}
\end{spacing}

\pagebreak

\end{document}